\documentclass[prl,twocolumn,amsmath,amssymb,floatfix]{revtex4-1}
\usepackage{graphicx}
\usepackage{dcolumn}
\newcolumntype{d}[1]{D{.}{.}{#1}}
\usepackage{longtable}

\usepackage{epstopdf}

\newcommand{\alo}{$\alpha$-Al$_2$O$_3$\ }
\newcommand{\gao}{$\beta$-Ga$_2$O$_3$\ }
\newcommand{\ino}{$\beta$-In$_2$O$_3$\ }

\begin{document}

\title{First principles study of thermal conductivity of In$_2$O$_3$ in relation to
 Al$_2$O$_3$, Ga$_2$O$_3$, and KTaO$_3$}

\author{Alaska Subedi} 

\affiliation{CPHT, CNRS, Ecole Polytechnique, IP Paris, F-91128
  Palaiseau, France} 
\affiliation{Coll\`ege de France, 11 place
  Marcelin Berthelot, 75005 Paris, France}

\date{\today}

\begin{abstract}

I use first principles calculations to investigate the thermal conductivity of $\beta$-In$_2$O$_3$ and
compare the results with that of $\alpha$-Al$_2$O$_3$, $\beta$-Ga$_2$O$_3$, and KTaO$_3$. The calculated
thermal conductivity of $\beta$-In$_2$O$_3$  agrees well with the experimental data obtain recently, which
found that the low-temperature thermal conductivity in this material can reach values above 1000 W/mK. 
I find that the calculated thermal conductivity of $\beta$-Ga$_2$O$_3$ is larger than
that of $\beta$-In$_2$O$_3$ at all temperatures, which implies that $\beta$-Ga$_2$O$_3$ should also exhibit
high values of thermal conductivity at low temperatures.  The thermal conductivity of KTaO$_3$ calculated 
ignoring the temperature-dependent phonon softening of low-frequency modes give high-temperature values similar
that of $\beta$-Ga$_2$O$_3$.  However, the calculated thermal conductivity of KTaO$_3$ does not 
increase as steeply as that of the binary compounds at low temperatures, which results in KTaO$_3$ having the 
lowest low-temperature thermal conductivity despite having acoustic phonon velocities larger than that of
$\beta$-Ga$_2$O$_3$ and $\beta$-In$_2$O$_3$.  I attribute this to the fact that the acoustic phonon 
velocities at low frequencies in KTaO$_3$ is less uniformly distributed because its acoustic phonon branches 
are more dispersive compared to the binary oxides, which causes enhanced momentum loss even during the normal 
phonon-phonon scattering processes.  I also calculate thermal diffusivity using the theoretically obtained 
thermal conductivity and heat capacit and find that all four materials exhibit the expected $T^{-1}$ behavior at 
high temperatures.  Additionally, the calculated ratio of the average phonon scattering time to Planckian 
time is larger than the lower bound of 1 that has been observed empirically in numerous other materials.

\end{abstract}


\maketitle

\section{Introduction}

\ino\ is a technologically important material due to its use as a transparent
conductor when doped with tin \cite{ellm12,bier15}.  In addition, it has been viewed 
as a potential candidate
for use in high-power devices along with other binary oxides \alo and \gao
from the same column of the periodic table because of its wide band gap.  A limiting 
factor in high-power applications is the ability to dissipate heat that is generated in 
the devices.  The thermal conductivity of single-crystal \alo and \gao have been studied 
both experimentally and theoretically \cite{slac62,cahi88,dong18,gala14,guo15,hand15,jian18,vllo08,sant15,yan18,mu19},
 which show that these materials are worse conductors
of heat than silicon with the room-temperature (300 K) values of $\sim$35, $\sim$20 and 
$\sim$150 W/mK \cite{krem04}, respectively.

In contrast, thermal properties of single-crystal \ino have received relatively little 
attention due to difficulty in growing them.  Recently, Galazka \textit{et al.}\ have
developed a method to grow cm$^3$ size single crystals of \ino \cite{gala11,gala14b}, and
their thermal conductivity from 20 to 300 K has been measured by Xu \textit{et al.} \cite{xu21}.
They find room-temperature (300 K) thermal conductivity of 15 W/mK, which is slightly worse 
than that of \alo and \gao\!\!.  Surprisingly, they also find that the thermal conductivity 
of \ino increases rapidly as the temperature is lowered, with a peak value as high as 5000 W/mK
at 20 K that is surpassed by few other insulators such as \alo \cite{slac62} and diamond \cite{onn92}.
Remarkably, 
experiments on \gao show a peak thermal conductivity of only 530 W/mK at 25 K \cite{hand15}. 
These results motive a theoretical investigation of the thermal conductivity of \ino so that a comparative
study of the microscopic mechanism of thermal transport in these binary oxides can be performed.

Behnia, who was part of the group that measured the thermal conductivity of \ino \cite{xu21}, 
asked me: i) why the thermal conductivity of \ino peaks at such a high value, and ii) why 
ternary oxides do not show similar high peaks at low temperatures.  In this paper, I 
attempt to answer his questions using first principles calculations of phonon dispersions 
and three-phonon scattering interactions, which I incorporate in the solution of the Boltzmann
transport equation (BTE) to obtain the lattice thermal conductivity. In addition to the binary 
oxides \alo, \gao and \ino, I also study the simplest ternary oxide KTaO$_3$ without 
incorporating its phonon softening effects to investigate the consequence of having an extra 
atomic species on the low-temperature thermal conductivity.  I find the thermal conductivity 
of \ino at 300 K is 10 W/mK, and it increases to 1.2$\times$10$^3$ W/mK at 30 K (the lowest 
temperature that I could solve the BTE).  These calculated values are
lower than the measured value of 15 W/mK \cite{xu21}, but they capture the three orders 
of magnitude increase in thermal conductivity as the temperature is lowered.  I find that the 
thermal conductivity of ternary KTaO$_3$ rises more slowly than the binary oxides.  I attribute
this to the fact that velocities of the acoustic phonon branches are distributed more evenly
in the binary oxides compared to the ternary KTaO$_3$, which should cause relatively less
loss in momentum during the three-phonon scattering processes in the binary oxides relative 
to KTaO$_3$.

\section{Crystal Structures and Computational Approach}

\begin{figure}

  \includegraphics[width=0.45\columnwidth]{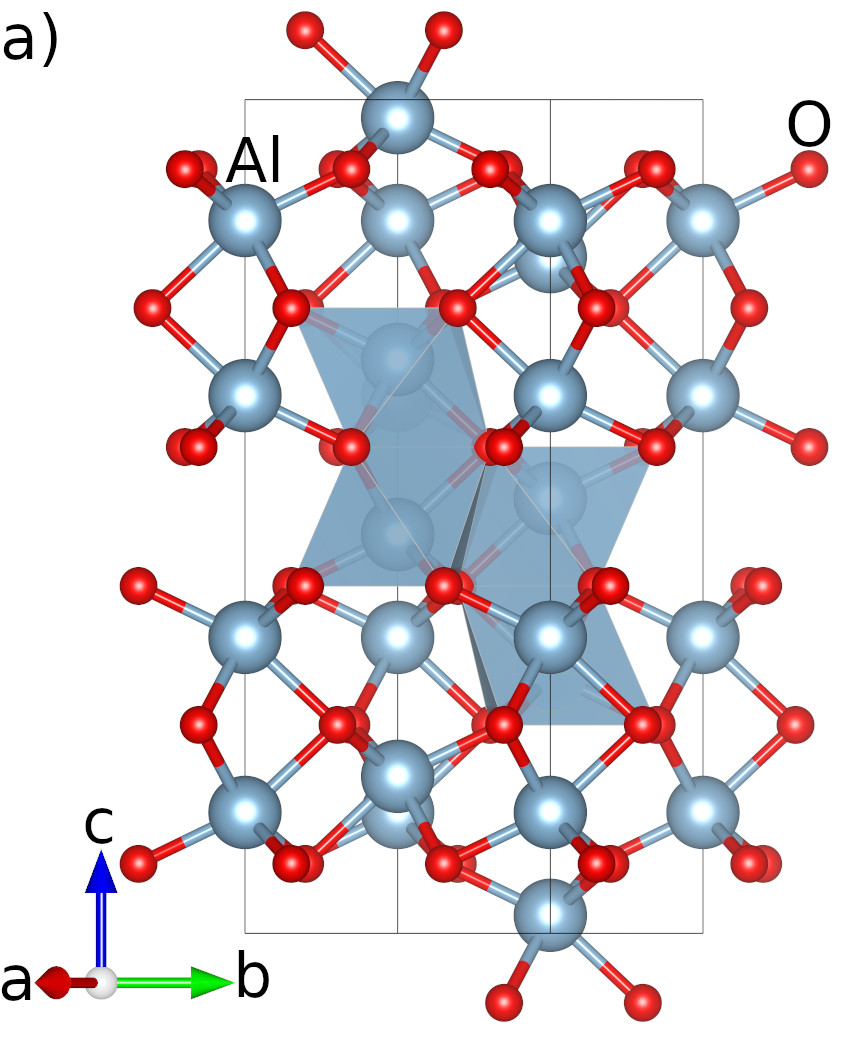}
  \includegraphics[width=0.4\columnwidth]{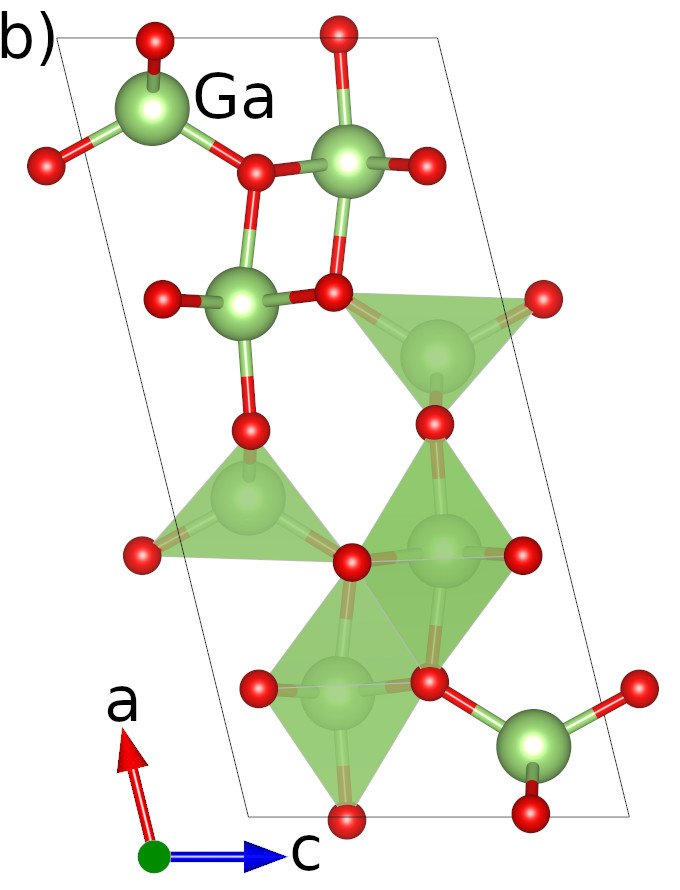}
  
  \includegraphics[width=0.45\columnwidth]{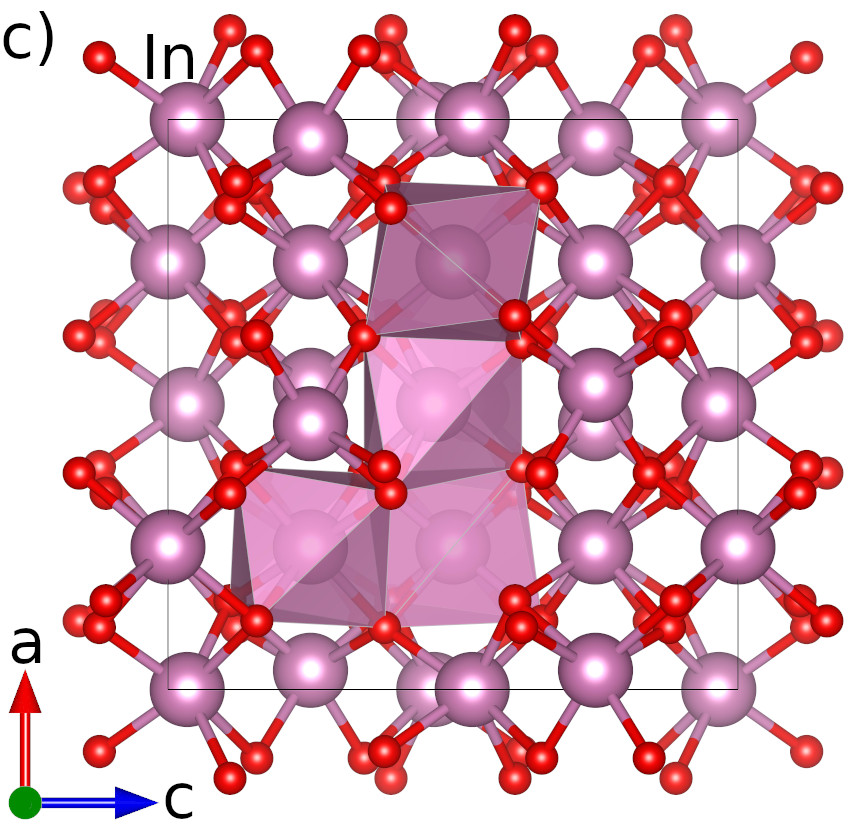}
  \includegraphics[width=0.44\columnwidth]{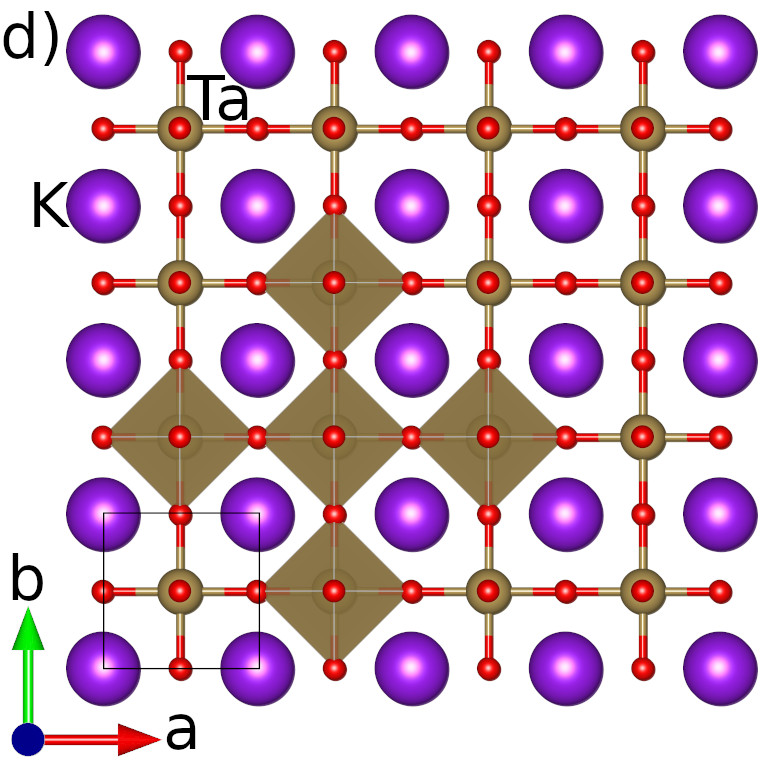}
  \caption{Crystal structures of a) $\alpha$-Al$_2$O$_3$, b) $\beta$-Ga$_2$O$_3$, 
  c) $\beta$-In$_2$O$_3$, and d) cubic KTaO$_3$. Some polyhedra are drawn to
  illustrate how they are connected to each other.}
  \label{fig:structs}
\end{figure}

The three binary $X_2$O$_3$ ($X$ = Al, Ga and In) compounds and the ternary KTaO$_3$ 
all occur with different crystal structures, which are shown in Fig.~\ref{fig:structs}.
$\alpha$-Al$_2$O$_3$ has a layered trigaonal structure with the space group 
$R\overline{3}c$ containing 10 atoms in the primitive unit cell.
Each layer consists of edge-shared AlO$_6$ octahedra arranged in a honeycomb pattern,
and the layers are stacked such that each octahedron shares edges with the
layers above and below it.
$\beta$-Ga$_2$O$_3$ has a highly anisotropic monoclinic structure with the
space group $C2/m$, and its primitive unit cell also contains 10 atoms. There are 
two nonequivalent Ga sites with tetrahedral and octahedral oxygen coordinations, respectively.  
The GaO$_4$ terahdra share their corners with GaO$_6$ octahedra, and each GaO$_6$
octahedron additionally has four edge-shared GaO$_6$ octahedra as neighbors. 
$\beta$-In$_2$O$_3$ occurs in the cubic bixbyite structure with the space group 
$Ia\overline{3}$ and has 40 atoms in its primitive unit cell. The structure 
consists of two inequivalent InO$_6$ octahedra, one undistorted and another highly
distorted, and each of them has six edge-shared and six corner-shared neighbors.
KTaO$_3$ has the 
simplest structure among all four compounds being studied in this work, 
occuring in cubic perovskite structure with the space group $Pm\overline{3}m$. 
Its primitive unit cell has five atoms, with the Ta ions situated inside 
corner-shared oxygen octahedra.

I calculated the second- and third-order interatomic force constants (IFCs) of these four
materials from first principles using density functional theory.  These were used 
to construct the linearized BTE
\begin{eqnarray}
\label{eq:lbte}
\mathbf{F}_{\lambda} & = & \tau_{\lambda} (\mathbf{v}_{\tau} + \mathbf{\Delta}_{\lambda}),
\end{eqnarray}
where $\tau_{\lambda}$ is the relaxation time of the phonon mode $\lambda$ obtained using
perturbation theory, $\mathbf{v}_{\tau}$ is the mode's group velocity, and 
$\mathbf{\Delta}_{\lambda}$ is the correction to the population of the mode $\lambda$ 
obtained from the simple relaxation time approach. Eq.~\ref{eq:lbte} was solved iteratively 
using the {\sc shengbte} package \cite{shengbte} to obtain the lattice thermal conductivity tensor 
$\kappa^{\alpha\beta}$, which is given by the expression
\begin{eqnarray}
    \kappa^{\alpha\beta} = \sum_{\lambda} C_{\lambda} v_{\lambda}^{\alpha} F_{\lambda}^{\beta}.
\end{eqnarray}
Here, $C_{\lambda}$ is the contribution of mode $\lambda$ to the specific heat. In addition to
the $\kappa^{\alpha\beta}$, {\sc shengbte} was also used to compute group velocities and 
Gr\"uneisen parameter $\gamma$ \cite{gruneisen} of each phonon mode, phonon contribution to heat capacity 
per mole $C$, and the weighted three-phonon scattering phase space $W$ \cite{wp3}. 

The density functional calculations were performed using the plane-wave
projector-augmented-wave method within the local density approximation (LDA), as 
implemented in the {\sc vasp} software package \cite{vasp}.  A plane-wave cutoff of 520 eV for
basis set expansion was used in all calculations.  I used potentials
in which the $3s$ and $3p$ (Al), $3d$, $4s$, and $4p$ (Ga), $4d$, $5s$, and $5p$ (In),
and $2s$ and $2p$ (O) were treated explicitly as valence electrons.  The interatomic
potentials were obtained using fully relaxed structures minimizing both the 
stresses and forces.  The Brillouin zone integration in the structural relaxation
calculations were performed via $k$-point meshes of $9\times9\times9$ ($\alpha$-Al$_2$O$_3$),
$12\times12\times6$ ($\beta$-Ga$_2$O$_3$), $4\times4\times4$ ($\beta$-In$_2$O$_3$), and $12\times12\times12$ 
(KTaO$_3$). The fully-relaxed lattice parameters are $a$ = 5.096 \AA\ and 
$\alpha$ = 55.36$^\circ$ ($\alpha$-Al$_2$O$_3$), $a$ = 12.160, $b$ = 3.025, $c$ = 5.768 \AA, and 
$\beta$ = 103.75$^\circ$ ($\beta$-Ga$_2$O$_3$),  $a$ = 10.705 \AA\ ($\beta$-In$_2$O$_3$),
and $a$ = 3.960 \AA\ (KTaO$_3$).  For \gao, there is an ambiguity in the choice of primitive
cell lattice parameters and the Brillouin zone.  I chose the convention used in Ref.~\cite{peel15}.

The harmonic second-order IFCs and phonon dispersions were  obtained using the 
real-space supercell-based frozen phonon method, as implemented in the {\sc phonopy} 
package \cite{phonopy}.  The third-order IFCs were also obtained using the real-space supercell 
approach, as implemented in the {\sc thirdorder.py} program of the {\sc shengbte} package. The supercells and $k$-point
meshes used in the calculations of the IFCs, as well as the phonon momentum-space $q$-point
meshes used in the solution of the Boltzmann transport equation for the four compounds 
are given in Table \ref{tab:grids}.

 \begin{table}
  \caption{\label{tab:grids} Dimensions of the supercells and $k$-meshes used in the calculations of
  second- and third-order IFCs, as well as that of the momentum-space $q$-mesh used in the solution 
  of the BTE for all four compounds studied in this work.}
  \begin{ruledtabular}
     \begin{tabular}{l c c c c c }
        & \multicolumn{2}{c}{second-order IFCs} & \multicolumn{2}{c}{third-order IFCs} & BTE \\
        & supercell & $k$-mesh  & supercell & $k$-mesh & $q$-mesh \\
       \hline
       \alo     & 4$\times$4$\times$4 & 3$\times$3$\times$3 & 3$\times$3$\times$3 & 3$\times$3$\times$3 & 14$\times$14$\times$14 \\
       \gao     & 4$\times$4$\times$4 & 3$\times$3$\times$2 & 3$\times$3$\times$3 & 3$\times$3$\times$2 & 12$\times$14$\times$12 \\
       \ino     & 2$\times$2$\times$2 & 2$\times$2$\times$2 & 2$\times$2$\times$2 & 2$\times$2$\times$2 & 9$\times$9$\times$9 \\
       KTaO$_3$ & 4$\times$4$\times$4 & 3$\times$3$\times$3 & 4$\times$4$\times$4 & 4$\times$4$\times$4 & 18$\times$18$\times$18 
    \end{tabular}
  \end{ruledtabular}
\end{table}

\section{Results and discussion}

The calculated effective thermal conductivities ($\kappa = \textrm{Tr}(\kappa^{\alpha\beta})/3$) 
of $\alpha$-Al$_2$O$_3$, $\beta$-Ga$_2$O$_3$, $\beta$-In$_2$O$_3$, and KTaO$_3$ using the fully 
relaxed LDA structures are shown in Fig.~1 along with the respective experimental data. Additionally,
the calculated and experimental values at 100 and 300 K are given in Table \ref{tab:kappa}. My calculated
thermal conductivity for \alo is in excellent agreement with a previous theoretical study \cite{dong18}.
The calculated and experimental \cite{cahi88} $\kappa$ for \alo agree very well at 300 K, but the experimental
$\kappa$ at 100 K is larger than the calculated value by 22\%.

There are three previous calculations of 
$\beta$-Ga$_2$O$_3$'s \cite{sant15,yan18,mu19}.  My results agree more with that of Ref.~\cite{sant15} than the
other two, presumably because we both use LDA while the other two studies use the generalized gradient 
approximation. The calculated $\kappa$ from the present study shows excellent agreement at 100 K compared
to the experimental data from Ref.~\cite{guo15}, but it overestimates the experimental value at 300 K by 30\%. 

\begin{figure}
 \includegraphics[width=\columnwidth]{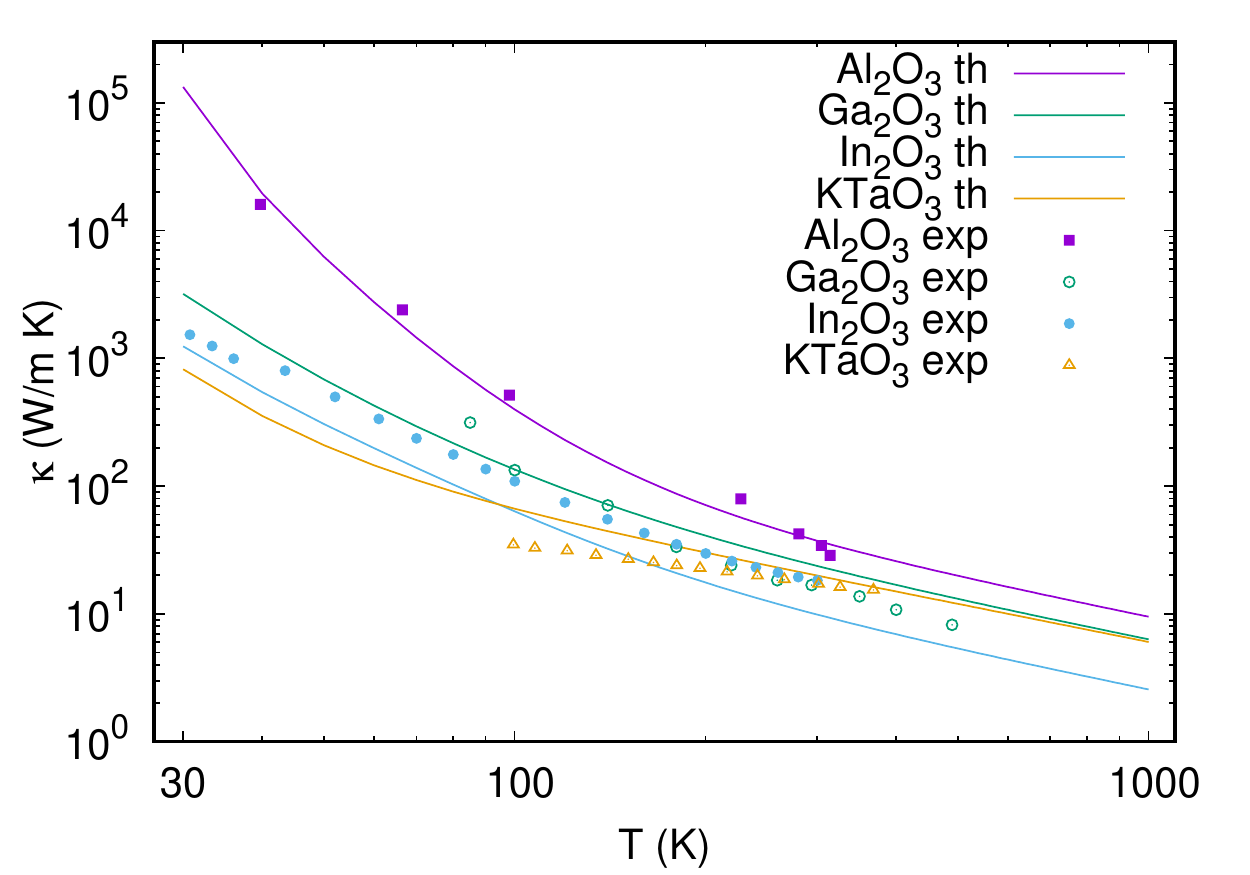}
  \caption{Calculated effective thermal conductivity ($\kappa = \textrm{Tr}(\kappa^{\alpha\beta})/3$) of
  $\alpha$-Al$_2$O$_3$, $\beta$-Ga$_2$O$_3$, $\beta$-In$_2$O$_3$, and KTaO$_3$ using the fully relaxed LDA crystal structures (solid lines). The experimental values are extracted from Refs.~\cite{cahi88,guo15,xu21,mart18}
  for $\alpha$-Al$_2$O$_3$, $\beta$-Ga$_2$O$_3$, $\beta$-In$_2$O$_3$, and KTaO$_3$, respectively.}
  \label{fig:kappa}
\end{figure}

For $\beta$-In$_2$O$_3$, the experimental $\kappa$ from Xu \textit{et al.}\ \cite{xu21} for their most conductive sample that was annealed in the presence of H$_2$  (shown in Fig.~\ref{fig:kappa}) is systematically larger than the calculated one from the present study at every temperature. They find that the thermal conductivity of samples 
annealed in O$_2$, and additionally in H$_2$, peaks to values above a thousand W/mK at temperatures below 30 K, but
as-grown and air-annealed samples show peaks that reach only up to 500 W/mK.  My results show that the intrinsic
$\kappa$ of \ino reaches as high as 1200 W/mK at 30 K, which indicates that the high thermal conductivity in \ino
measured in Ref.~\cite{xu21} reflects the intrinsic transport property of the material.  Nevertheless, the fact that
my calculations underestimate the experimental $\kappa$ suggests that impurities may still play a role in enhancing
the thermal conductivity of this material.  Anyway, the experimental thermal conductivity of all samples start to converge above 100 K.  The experimental $\kappa$ at 100 and 300 K
of \ino annealed in O$_2$ is given in Table \ref{tab:kappa}, and they are larger than the theoretical
values by around 33\%. Interestingly, my calculations show that the effective thermal conductivity of \gao is 
larger than that of \ino at every temperature.  This implies that \gao may show peak thermal conductivity of at least
a thousand W/mK at low temperatures. 

Lastly, the calculated $\kappa$ of KTaO$_3$ is in good agreement with the experimental values \cite{mart18} above 300 K. At 
lower temperatures, the theoretical values greatly overestimate the experiment, which is likely due to the neglect of
temperature dependent phonon softening of low frequency phonons that is experimentally observed in the material. 
A similar approximation in the calculation of thermal conductivity of this material was made in a previous theoretical
study \cite{fu18}, whose results are well reproduced here.  The disregard for phonon softening effects means that
the calculations does not explain the low-temperature thermal conductivity of KTaO$_3$.  Nevertheless, these 
calculations are helpful in illustrating the differences in the mechanism for thermal transport between the binary
and ternary oxides.

\begin{table}
  \caption{\label{tab:kappa} Calculated and experimental effective thermal conductivity  
  ($\kappa = \textrm{Tr}(\kappa^{\alpha\beta})/3$) of $\alpha$-Al$_2$O$_3$, $\beta$-Ga$_2$O$_3$,
   $\beta$-In$_2$O$_3$, and KTaO$_3$ at 100 and 300 K. The experimental values are extracted from 
   Refs.~\cite{cahi88,guo15,xu21,mart18}. The values are given in the units of W/mK.}
  \begin{ruledtabular}
     \begin{tabular}{l d{2.2} d{2.2} d{2.2} d{2.2}}
        & \multicolumn{2}{c}{100 K} &  \multicolumn{2}{c}{300 K}  \\ 
        & \multicolumn{1}{c}{$\kappa^{\textrm{th}}$} & \multicolumn{1}{c}{$\kappa^{\textrm{exp}}$}
        & \multicolumn{1}{c}{$\kappa^{\textrm{th}}$} & \multicolumn{1}{c}{$\kappa^{\textrm{exp}}$}\\
        \hline
        \alo     & 399.19 & 516 & 37.38 & 34 \\
        \gao     & 134.91 & 133 & 23.75 & 17 \\
        \ino     &  63.52 &  95 & 9.88  & 15 \\
        KTaO$_3$ &  66.60 &  35 & 19.95 & 17 \\
     \end{tabular}
  \end{ruledtabular}
\end{table}

Even though the three binary compounds occur in structures with different 
cation-oxygen bonding environment and network, they follow a decreasing trend 
in the magnitude of thermal conductivity as the atomic mass of their cation 
increases.  This is consistent with the conventional wisdom that low atomic
mass is conducive to high thermal conductivity \cite{slac73}. But a comparison 
with KTaO$_3$, 
which has a relatively simple cubic perovskite structure, shows that interatomic
properties can play an even larger role in determining the thermal conductivity.
The average atomic masses of the cations in In$_2$O$_3$ and KTaO$_3$ differ by 
only 4\%, but the calculated thermal conductivity at 400 K of KTaO$_3$ is closer 
to that of Ga$_2$O$_3$, whose cation atomic weight is lower by 37\%.  Therefore,
it is instructive to examine the microscopic interatomic quantities to identify
the main factors influencing the thermal conductivity of a material.

\begin{figure*}
  \includegraphics[width=\columnwidth]{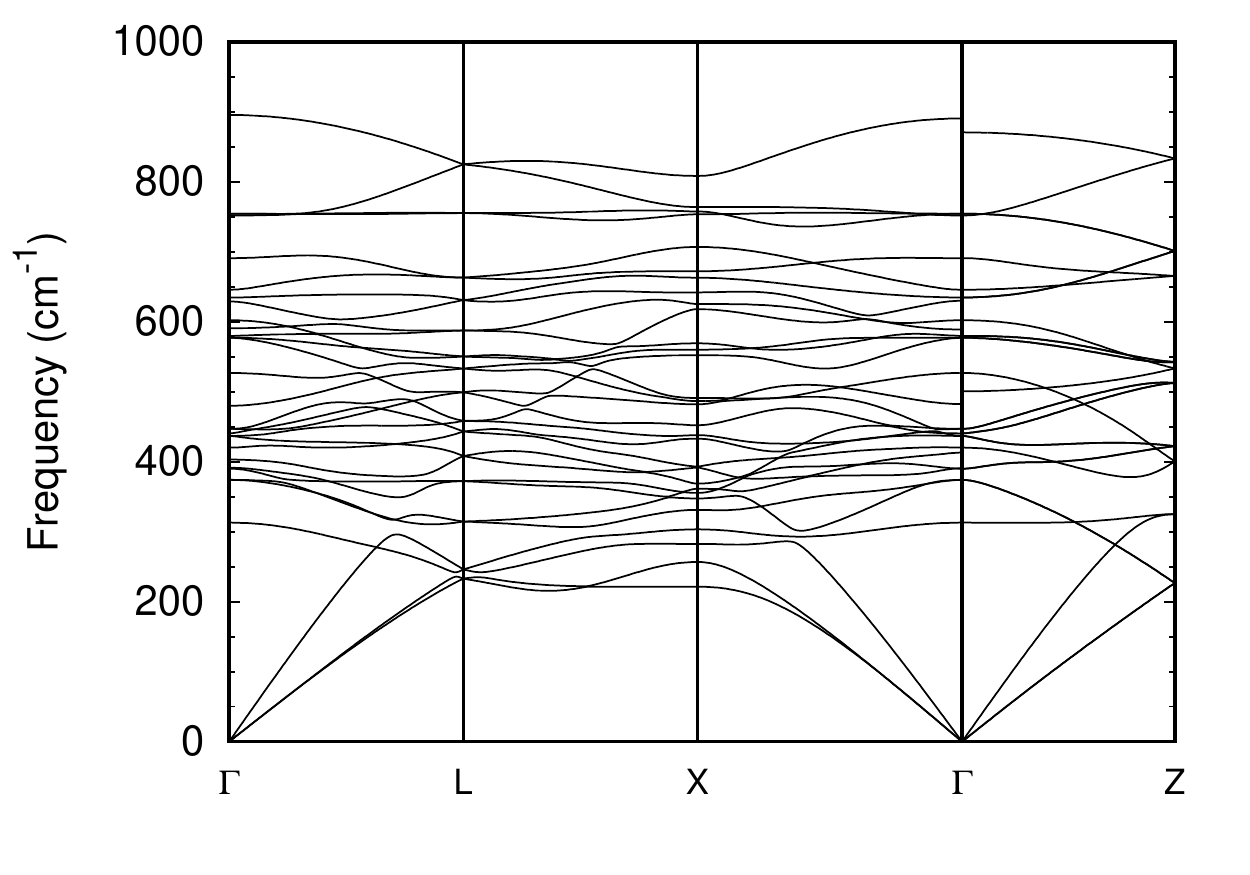}
  \includegraphics[width=\columnwidth]{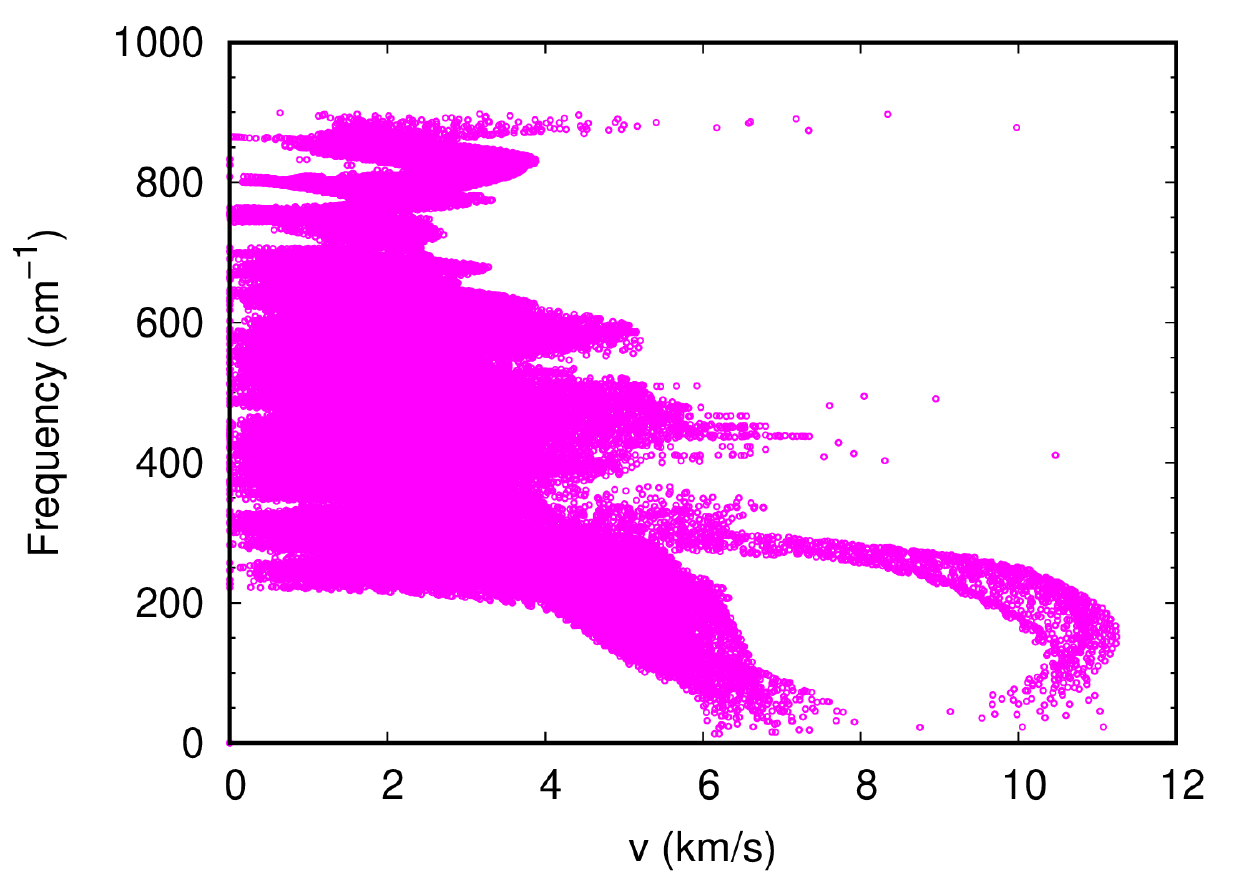}
  \includegraphics[width=\columnwidth]{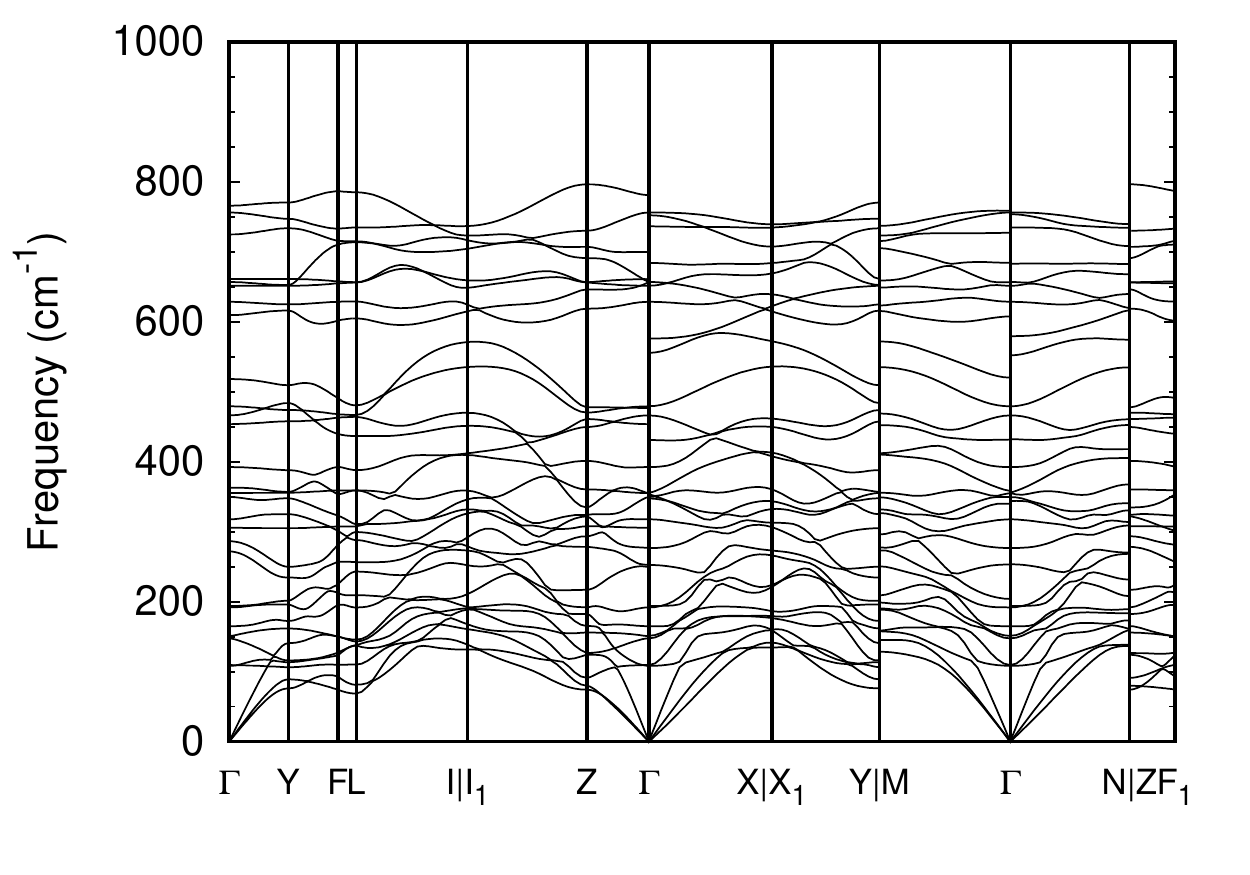}
  \includegraphics[width=\columnwidth]{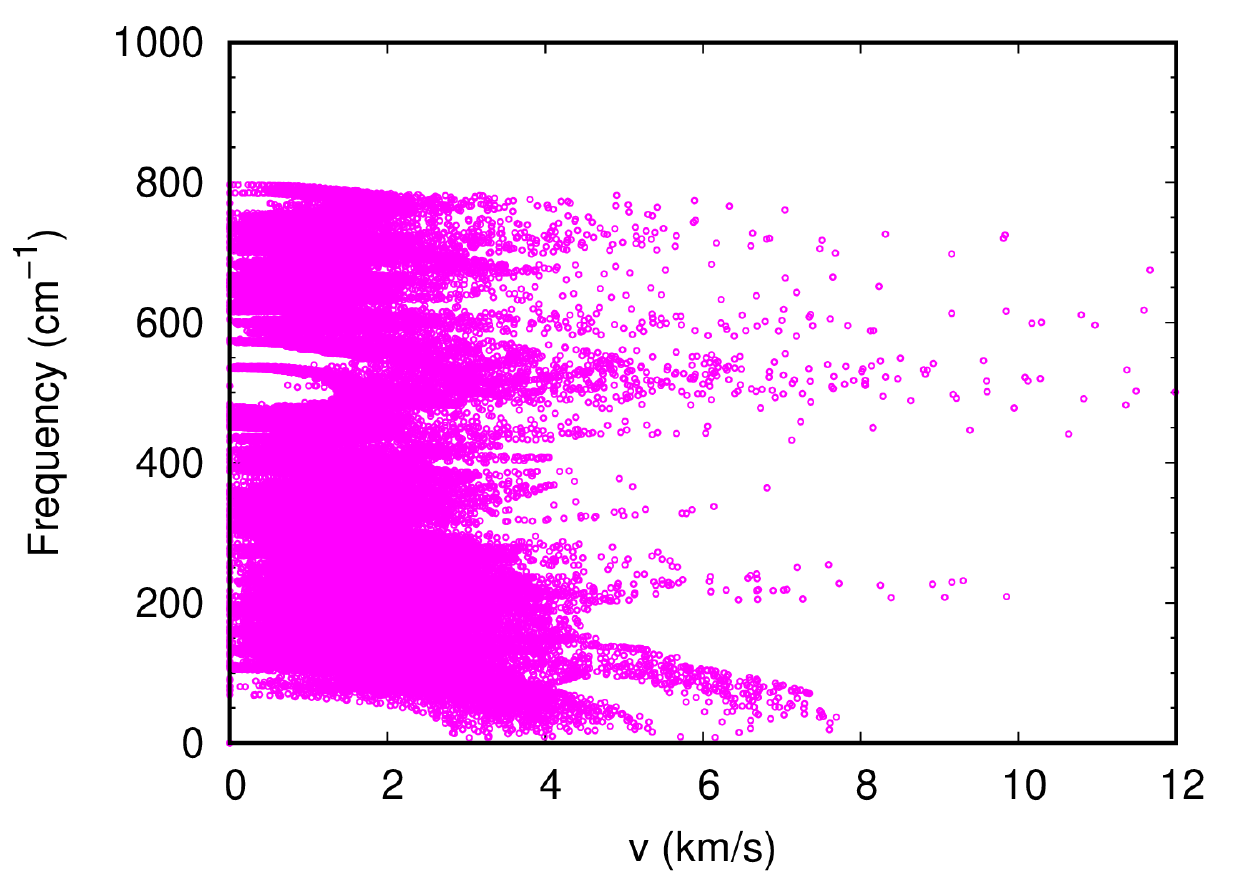}
  \includegraphics[width=\columnwidth]{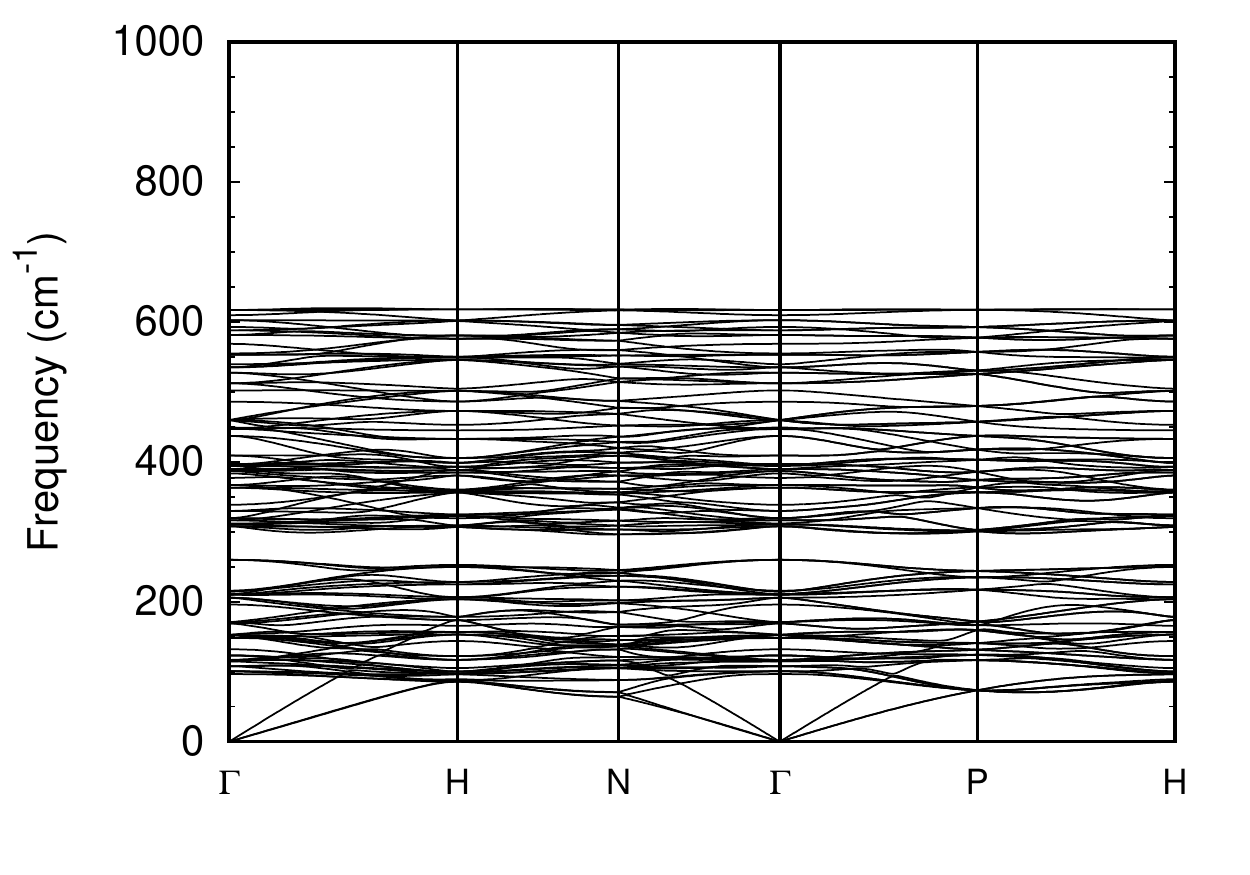}
  \includegraphics[width=\columnwidth]{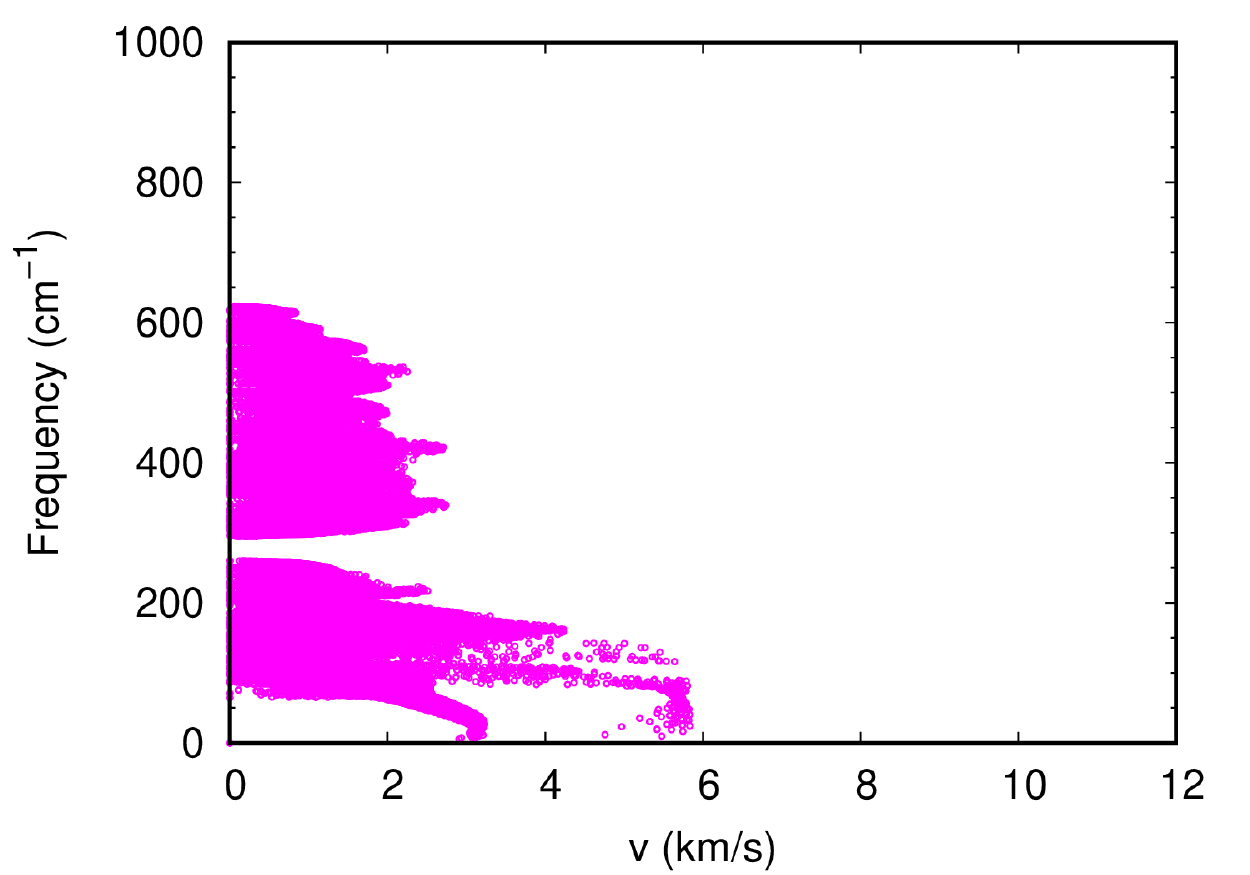}
  \includegraphics[width=\columnwidth]{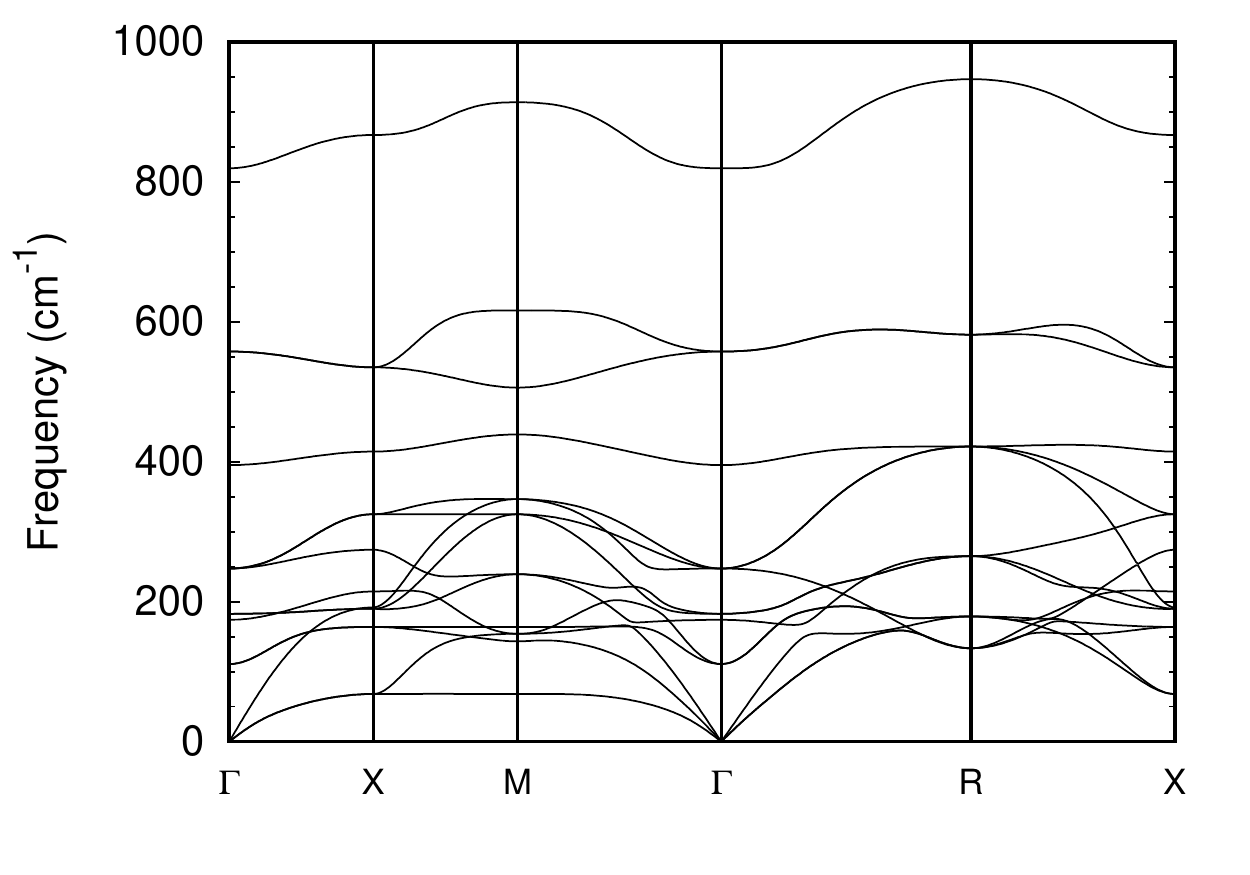}
  \includegraphics[width=\columnwidth]{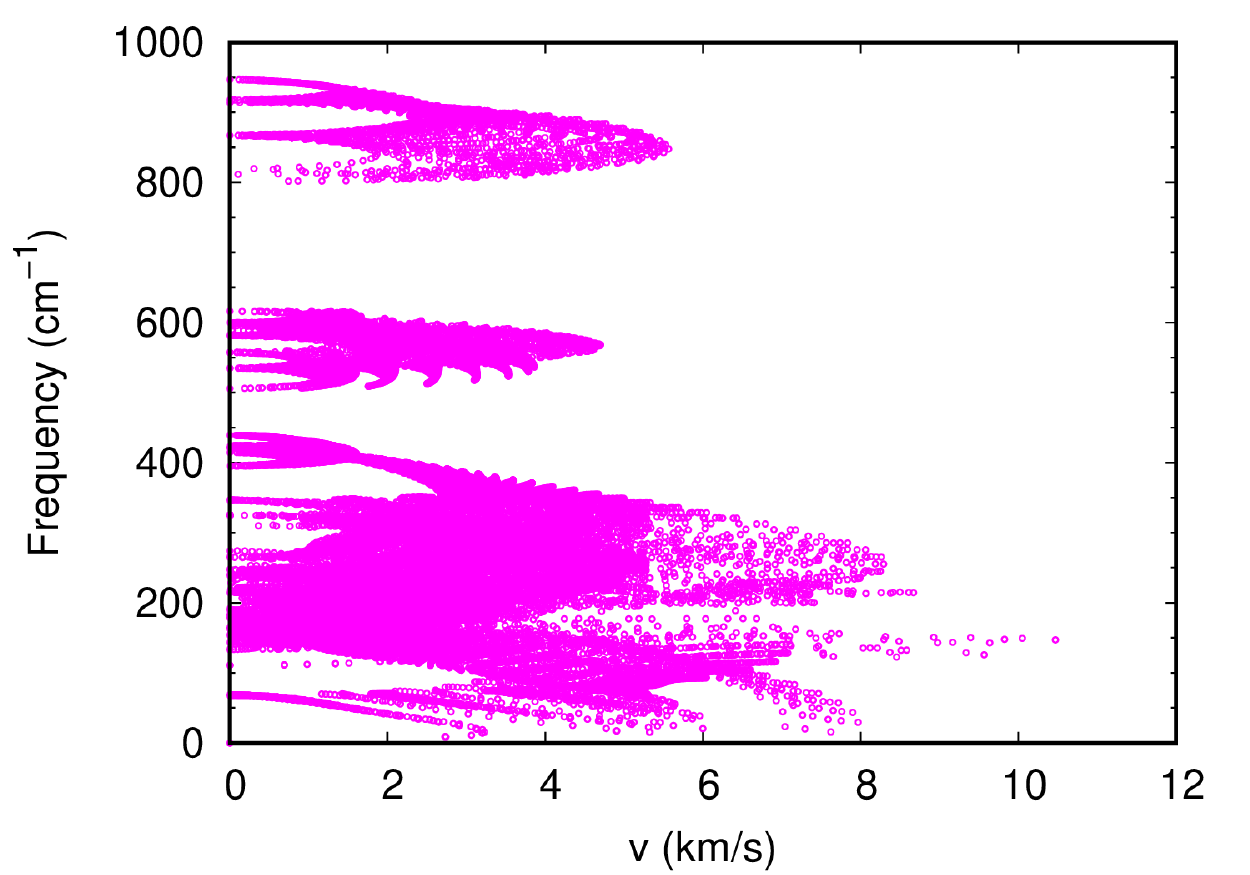}
  \caption{(Left) Calculated phonon dispersions. (Right) The respective phonon
  velocities. The compounds are $\alpha$-Al$_2$O$_3$, $\beta$-Ga$_2$O$_3$, 
   $\beta$-In$_2$O$_3$ and KTaO$_3$ from top to bottom, respectively. }
  \label{fig:disp-vel}
\end{figure*}

\begin{figure}
  \includegraphics[width=\columnwidth]{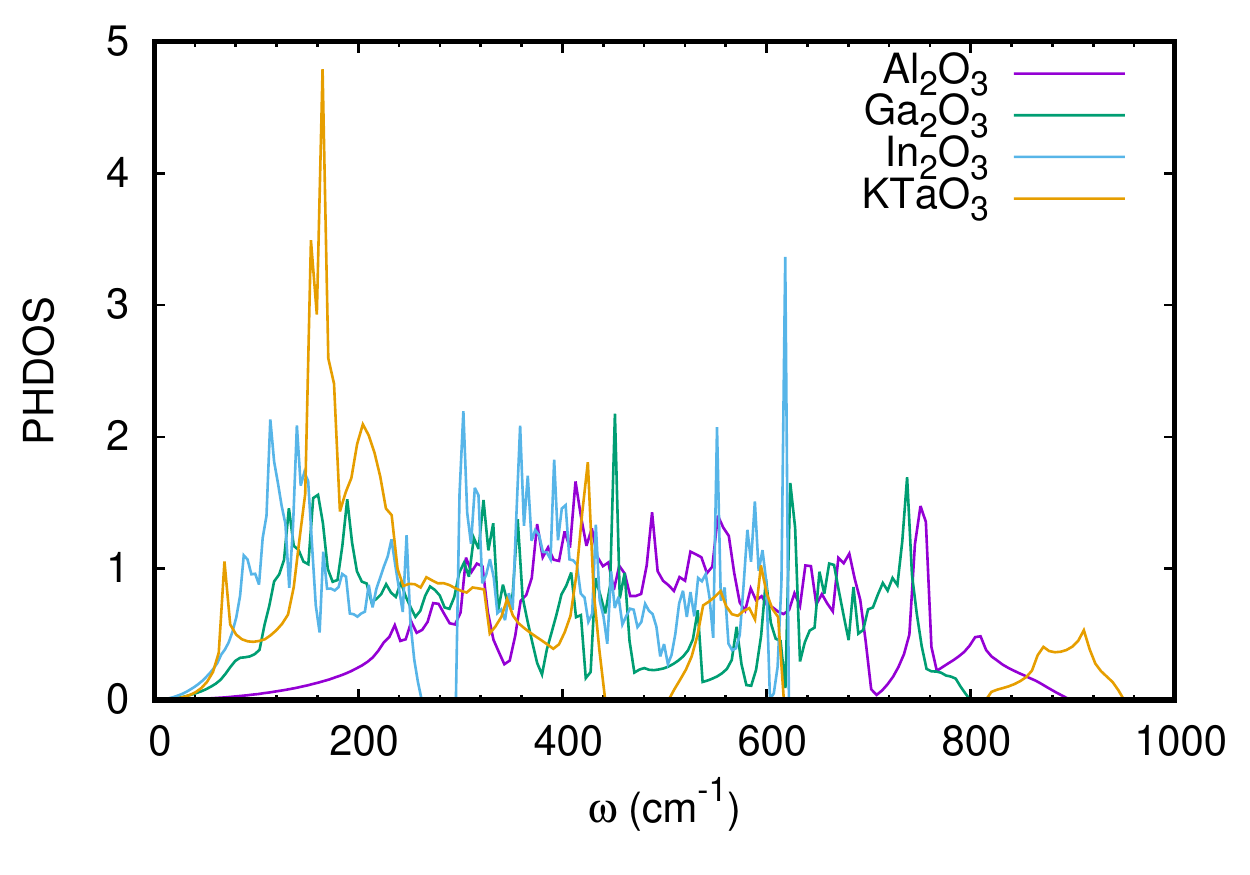}
  \caption{Calculated phonon density of states normalized to per formula unit (\textit{i.e.}\ five atoms).}
  \label{fig:phdos}
\end{figure}

The calculated phonon dispersions of all four materials being studied in this work
are shown in the left column of Fig.~\ref{fig:disp-vel}, and the respective phonon 
density of states (PHDOS) are 
shown in Fig.~\ref{fig:phdos}.  KTaO$_3$ has the most uncluttered phonon dispersions, 
reflecting its simple perovskite structure with only five atoms per primitive unit cell.
The intricacy of the dispersions increases as the complexity of the crystal structure 
increases from $\alpha$-Al$_2$O$_3$ to $\beta$-Ga$_2$O$_3$ to $\beta$-In$_2$O$_3$. 
It has been previously noted that complex crystal structures and phonon dispersions 
both lead to low values of thermal conductivity at high temperatures \cite{slac73,slac79}.
Indeed, these four compounds have five or more atoms per unit cell, and their
calculated thermal conductivity decrease strongly at high 
temperatures, becoming less than 10 W/mK at 1000 K.  Nevertheless, the thermal 
conductivity at 1000 K of KTaO$_3$ (6 W/mK) is appreciably smaller than that of 
Al$_2$O$_3$ (9.3 W/mK), which shows that having a more complex and dense phonon 
dispersions at high frequencies does not necessarily lead to lower thermal conductivity 
at high temperatures.

Since lattice thermal conductivity is proportional to the square of the group velocities of 
phonons, high acoustic phonon velocities should lead to large values of thermal conductivity,
especially at low temperatures where heat is mostly carried by long-wavelength acoustic phonons.
The right column of 
Fig.~\ref{fig:disp-vel} shows the norm of the velocity of each phonon mode as a function 
of frequency in the four materials.  $\alpha$-Al$_2$O$_3$ has the highest value of thermal
conductivity, and its acoustic phonons indeed have the highest velocities that reach up to 
11.3 km/s. However, KTaO$_3$ and \gao have similarly high values
of acoustic phonon velocities, but the calculated thermal conductivity of KTaO$_3$ is almost
four times lower than that of \gao in the low-temperature limit.  In fact, the acoustic phonon 
velocities of KTaO$_3$ are 
higher than that of $\beta$-In$_2$O$_3$, but the low-temperature thermal conductivity of KTaO$_3$
is lower. This suggests that merely having high acoustic 
phonon velocities does not necessarily imply high thermal conductivity in the low temperature 
regime.  

A closer look at Fig.~\ref{fig:disp-vel} shows that the phonon velocity distributions at low 
frequencies in the binary compounds are less dispersive than in KTaO$_3$.  
In \alo and \ino\!\!, the acoustic phonon velocities are bunched up in two narrow groups deriving 
from the respective transverse and longitudinal branches for phonons with frequencies less than
 50 cm$^{-1}$. The acoustic phonon velocity 
distribution is slightly wider in \gao, but the velocities still do not get lower 
than 2.2 km/s for the acoustic phonons below 50 cm$^{-1}$ in frequency. In contrast, the 
low-frequency velocity distribution in the same frequency window is more scattered in KTaO$_3$, and the phonon velocities get 
as low as 1.4 km/s. When acoustic phonons at low frequencies occur with a wide range of velocities, 
the loss of momentum when they collide with each other during heat transport is more enhanced 
even for normal scattering processes. Therefore, a more uniform distribution of phonon velocities with 
smaller magnitudes is 
likely to yield higher thermal conductivity than a phonon velocity distribution with high velocities 
that rapidly change as a function of 
frequency.

The phonon velocity distribution shown in Fig.~\ref{fig:disp-vel}(right) can also qualitatively 
account for why the high-temperature thermal conductivities of $\alpha$-Al$_2$O$_3$, 
$\beta$-Ga$_2$O$_3$ and KTaO$_3$ are larger than that of $\beta$-In$_2$O$_3$.  The optical phonon
branches of $\beta$-In$_2$O$_3$ only extend up to 625 cm$^{-1}$, and they have comparatively 
small magnitudes of velocities with values less than 2.8 km/s.  In contrast, the optical branches
of $\alpha$-Al$_2$O$_3$, $\beta$-Ga$_2$O$_3$ and KTaO$_3$ extend up to at least 800 cm$^{-1}$,
and some of these high-frequency optical phonons have velocity in excess of 4 km/s.  Therefore,
the differences in the phonon velocities of the high-frequency optical phonons of these materials
partially explains the differences in their high-temperature thermal conductivity. 

\begin{figure}
  \includegraphics[width=\columnwidth]{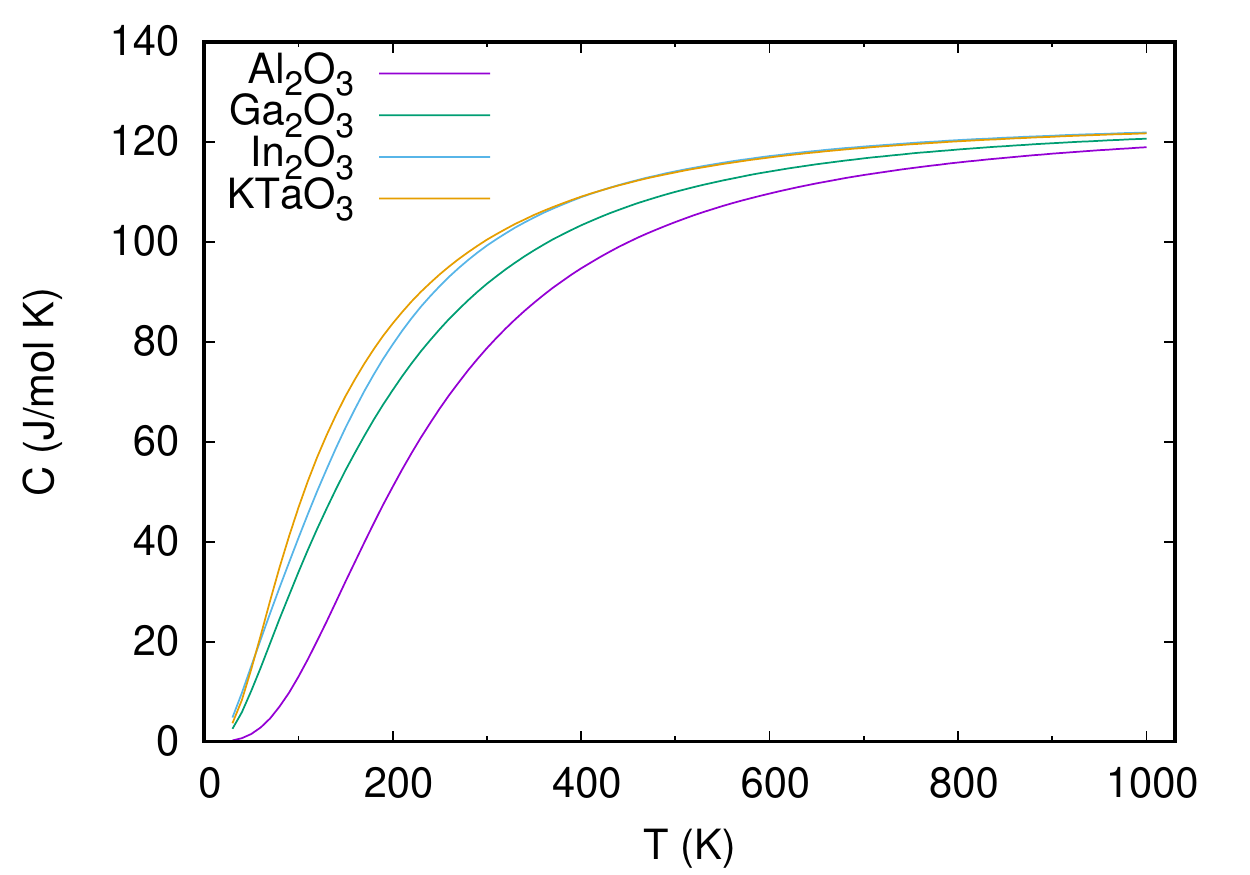}
  \caption{Calculated molar heat capacity as a function of temperature.}
  \label{fig:cv}
\end{figure}

Other microscopic quantities such as PHDOS, heat capacity, three-phonon scattering
phase space and Gr\"uneisen parameter also play a role in determining the lattice thermal conductivity 
of a material, but their relative influence have been debated. As one can see in Fig.~\ref{fig:phdos}, 
it is difficult to gain much understanding of the difference in the calculated thermal conductivities 
of the four compounds by analyzing their phonon density of states.  Both $\beta$-Ga$_2$O$_3$ and 
KTaO$_3$ have similar values of PHDOS up to 40 cm$^{-1}$, but their low-temperature thermal 
conductivities differ by more than a factor of two.  Above 300 cm$^{-1}$, the PHDOS of the four compounds 
overlap considerably, and there is no obvious structure in the PHDOS to account for the different values 
of the calculated thermal conductivities of the four materials at high temperatures.  Similarly, the 
differences in the calculated lattice heat capacity of the four materials shown in Fig.~\ref{fig:cv} also 
does not qualitatively account for the variance in the thermal conductivity at any given temperatures.
For example, $\alpha$-Al$_2$O$_3$ has the highest calculated thermal conductivity among the four compounds at all 
temperatures, but it has the lowest heat capacity and reaches saturation less rapidly reflecting its slowly 
increasing PHDOS.
   
\begin{figure}[h]
  \includegraphics[width=\columnwidth]{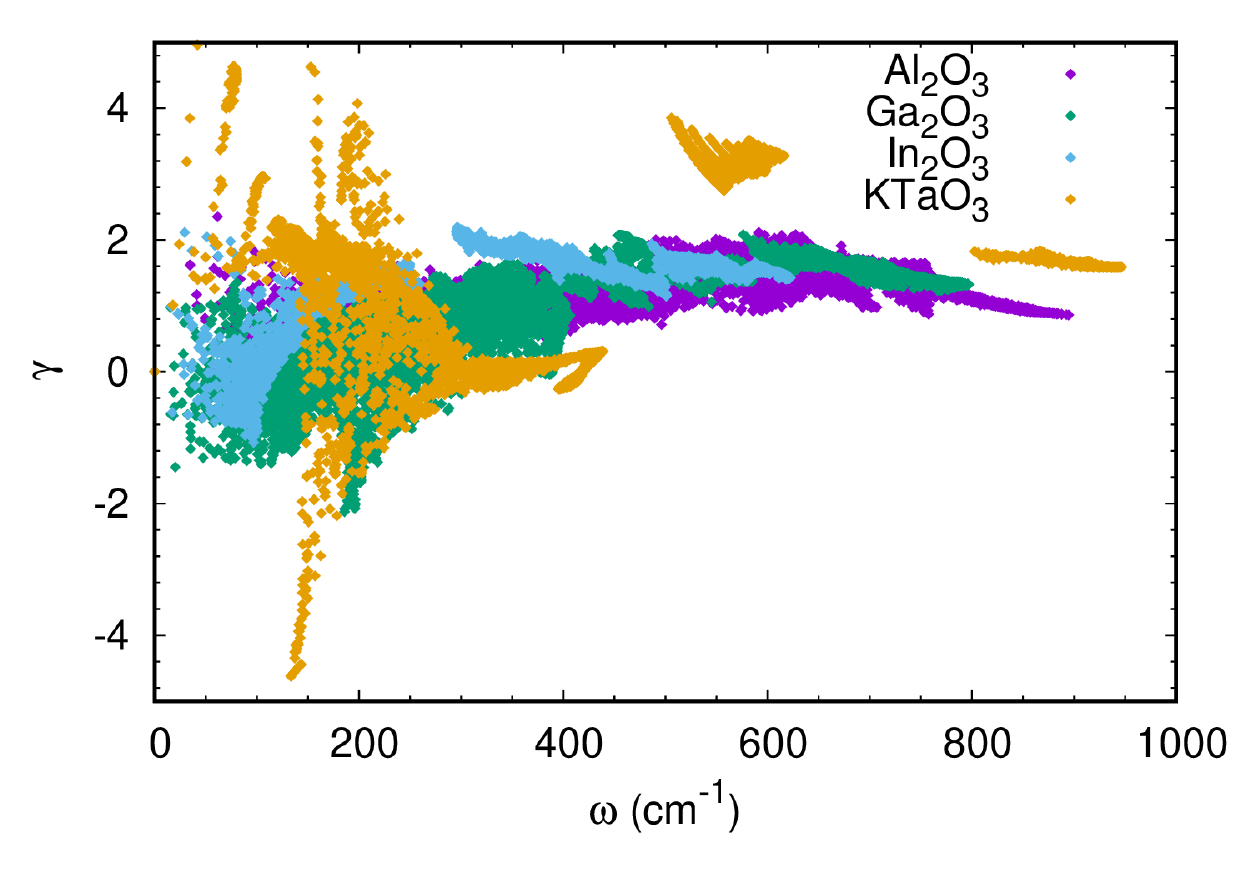}
  \caption{Calculated Gruneisen parameter for each phonon mode.}
  \label{fig:grun}
\end{figure}

The Gr\"uneisen parameters, which reflect the anharmonicity of phonons, of the four compounds are shown
Fig.~\ref{fig:grun}.  There is a big overlap of the calculated values throughout the frequency range of
the phonon spectrum. $\alpha$-Al$_2$O$_3$'s Gr\"uneisen parameters are positive, reflecting the fact that
its phonons harden upon volume contraction. $\beta$-Ga$_2$O$_3$ and $\beta$-In$_2$O$_3$ additionally have
low frequency phonons with negative Gr\"uneisen parameters.  The magnitude of the Gr\"uneisen parameters 
of all three binary compounds are relatively modest and similar, indicating that the variance in the 
calculated thermal conductivity of these compounds does not primarily derive from the anharmonicity of 
phonons.  However, the Gr\"uneisen parameters of KTaO$_3$ are much larger at low frequencies, which implies
the presence of stronger three-phonon interactions. This may be the reason why the thermal conductivity
of this compound increases more slowly as the temperature is decreased compared to the three binary
compounds.

\begin{figure}
  \includegraphics[width=\columnwidth]{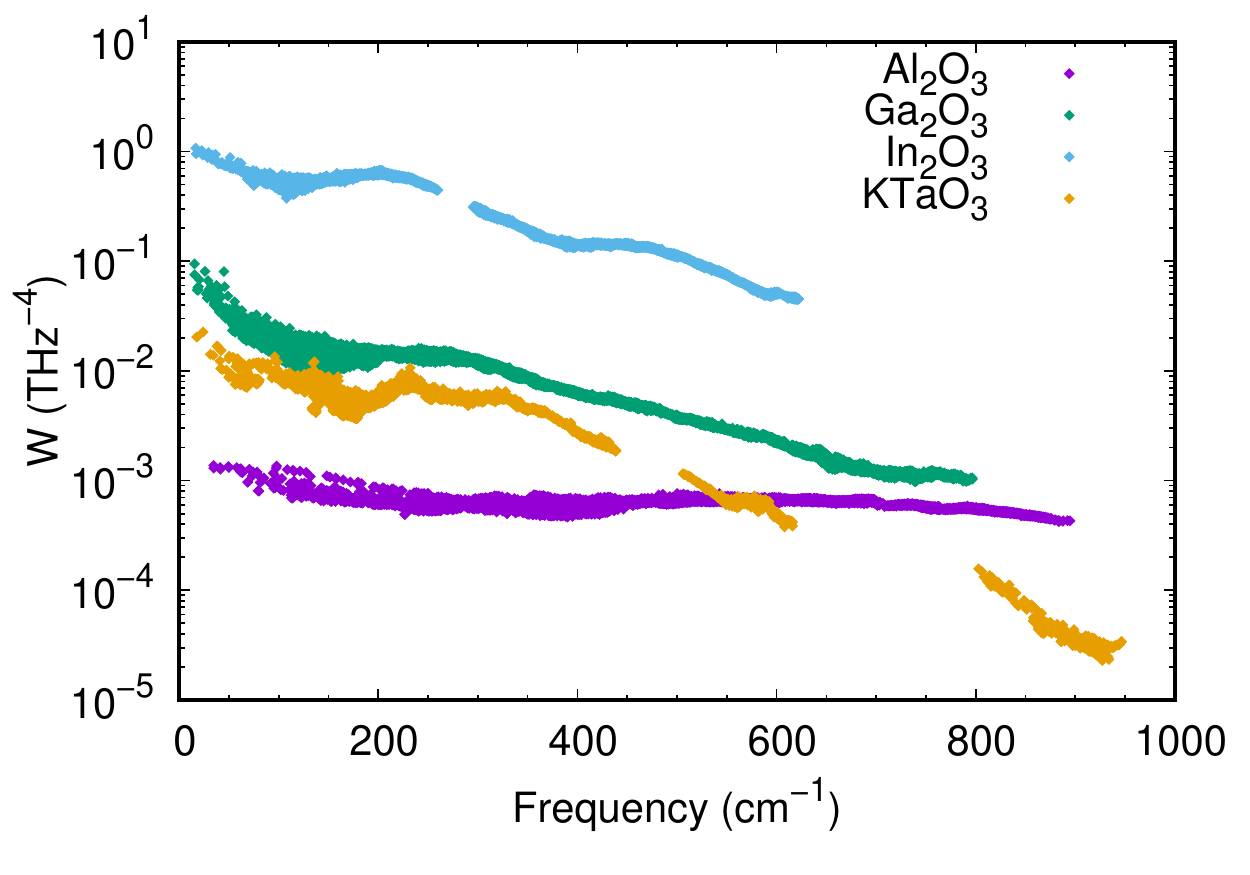}
  \caption{Calculated weighted three-phonon scattering phase space at 300 K.}
  \label{fig:wp3}
\end{figure}

The calculated weighted scattering phase space available for three-phonon absorption and emission 
processes of the
four compounds is shown in Fig.~\ref{fig:wp3}, and they are distinct for each compound.  For the three
binary compounds, the clear difference in their thermal conductivity is apparent in the scattering phase
space.  The scattering phase space increases by more than an order of magnitude below 300 cm$^{-1}$ each 
time the cation species moves to the lower row in the periodic table, which unambiguously demonstrates 
that the scattering phase space is inversely proportional to the calculated thermal conductivity in the 
three binary compounds.  However, the scattering phase space of the ternary compound KTaO$_3$ does not 
follow the same trend.  Its scattering phase space at high frequencies is lower by more than an order of 
magnitude than that of $\alpha$-Al$_2$O$_3$. But at low frequencies, its scattering phase space is larger 
by an order of magnitude.  This suggests that simply have extra elements in a compound can dramatically
increase the three-phonon scattering phase space at low frequencies.

 \begin{figure}
  \includegraphics[width=\columnwidth]{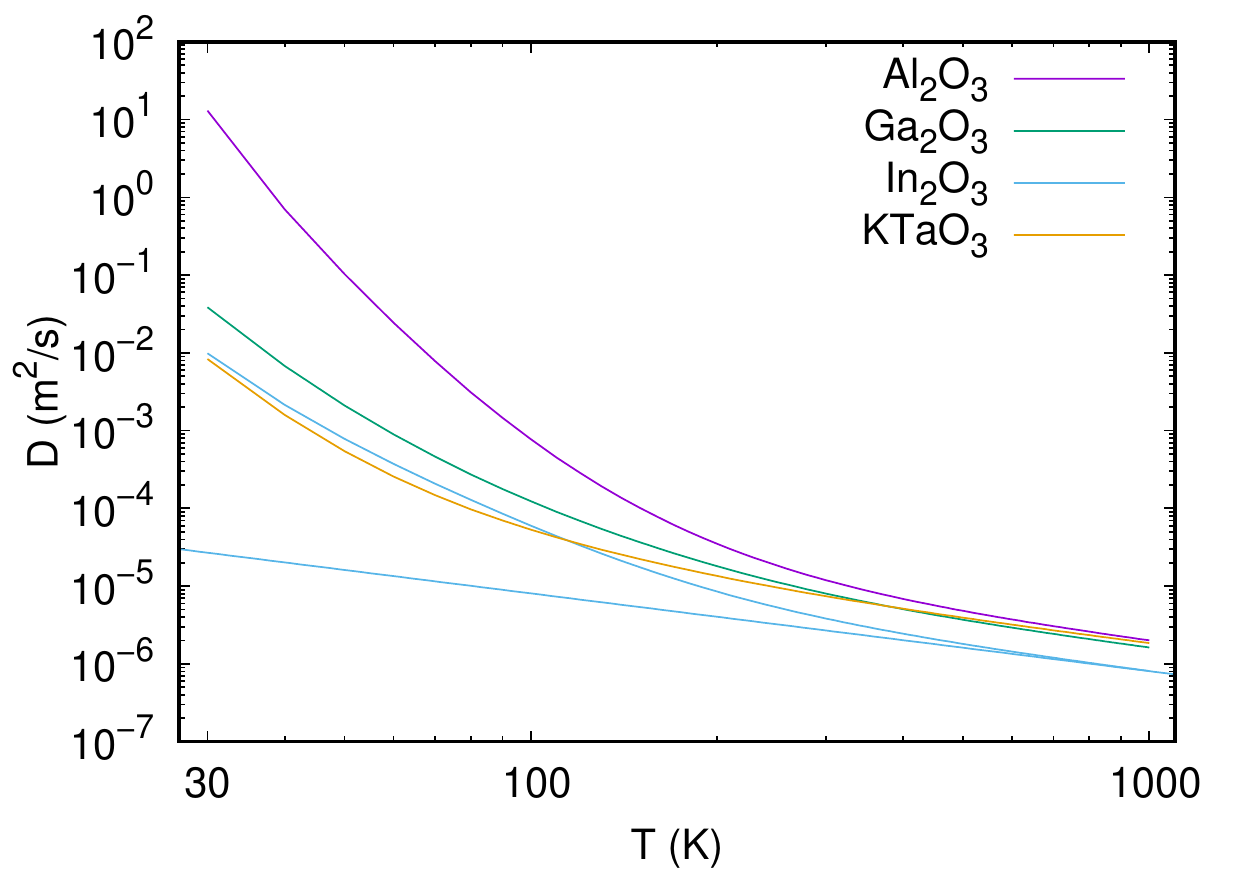}
  \caption{Calculated thermal diffusivity as a function of temperature.  The straight line is calculated
  using $v_s = 5.83$ km/s and $s = 3.11$ obtained for \ino at 1000 K, and it shows that the calculated
  diffusivity exhibits a $T^{-1}$ behavior at high temperatures.}
  \label{fig:diffu}
\end{figure}

\begin{table}
  \caption{\label{tab:s} The maximum sound velocity $v_s$ (km/s) and calculated thermal 
  diffusivities $D^{\textrm{th}}$ (10$^{-6}$ m$^2$/s) at 300 and 1000 K of the four materials studied
   in this work. The experimental diffusivity at room temperature is also given.}
  \begin{ruledtabular}
     \begin{tabular}{l d{1.2} d{2} d{1.2} d{1.2}}
        & & \multicolumn{2}{c}{300 K} &  \multicolumn{1}{c}{1000 K}  \\ 
        & \multicolumn{1}{c}{$v_s$} & \multicolumn{1}{c}{$D^{\textrm{th}}$} 
        & \multicolumn{1}{c}{$D^{\textrm{exp}}$} & \multicolumn{1}{c}{$D^{\textrm{th}}$}  \\
        \hline
        \alo     & 11.08 & 11.94 & 13 & 2.01\\
        \gao     & 7.69  & 8.04  & -  & 1.63\\
        \ino     & 5.83  & 3.8   & 7  & 0.81\\
        KTaO$_3$ & 7.62  & 7.43  & -   & 1.85\\
     \end{tabular}
  \end{ruledtabular}
\end{table}


Finally, Fig.~\ref{fig:diffu} shows the thermal diffusivity $D$ = $\frac{\kappa}{C'}$ of all four 
compounds calculated using the same data used in Figs.~\ref{fig:kappa} and \ref{fig:cv}. Here, 
$C'$ is the heat capacity per unit volume.  
The diffusivity approaches a behavior that is proportional to $T^{-1}$ in all four compounds at high 
temperatures, which is expected since the heat capacity saturates to a constant value according the 
Dulong-Petit law and the phonon-phonon scattering rate is proportional to $T^{-1}$ above the Debye 
temperature.  Although the high-temperature diffusivity has the same temperature dependence in all four 
compounds, there is some variance in the magnitude of their diffusivity.  Again, a look back at the 
velocity distribution plot in Fig.~\ref{fig:disp-vel} shows that larger phonon velocities at high 
frequencies lead to larger high-temperature diffusivity, with \ino having both low velocities and low 
diffusivity than the other three materials.  The values of high-temperature diffusivity of \alo, \gao, 
and KTaO$_3$ lie close, and the remaining differences in their relative values of $D$ in the high-$T$ 
regime seems to reflect the differences in their acoustic phonon velocities, as both the acoustic phonon
velocities and high-$T$ diffusivity decrease in magnitude going from \alo to KTaO$_3$ to \gao.

It has recently been pointed out empirically that there is a lower limit to thermal diffusivity in 
insulating crystals \cite{mart18,behn19}. The high-temperature diffusivity can be approximately expressed as  
$D = s v_s^2 \tau_p$, where $\tau_p = \frac{\hbar}{k_B T}$ is the Planckian time, $v_s$ is the 
sound velocity, and $s$ is the dimensionless parameter representing the ratio of 
the average phonon scattering time $\tau$ to $\tau_p$.  Experimental data on numerous insulators 
find a lower-bound $s > 1$ \cite{behn19,mous20}, which implies that phonon-phonon collisions cannot happen 
faster than the Planckian time. Using the calculated $v_s$ and $D^{\textrm{th}}$ at 300 K, I obtain for $s$
the values of 3.8 ($\alpha$-Al$_2$O$_3$), 5.3 ($\beta$-Ga$_2$O$_3$), 4.4 ($\beta$-In$_2$O$_3$), and 5.0 (KTaO$_3$).
For comparison, the experiments on \ino gives $s = 5.8$ at 300 K \cite{xu21}.



\section{Summary and Conclusions}

In summary, I have used first principles calculations to perform a comparative study of thermal conductivity
in $\alpha$-Al$_2$O$_3$, $\beta$-Ga$_2$O$_3$, $\beta$-In$_2$O$_3$ and KTaO$_3$, which was motivated by
a recent experimental observation of very high low-temperature $\kappa$ in \ino by Xu \textit{et al.} My 
results on \ino agrees well with the experimental study, indicating that the peak conductivity value of around
1000 W/mK at low temperatures is an intrinsic property of this material.  The calculated $\kappa$ of \gao is 
also in reasonable agreement with the available experimental data above 80 K.  Additionally, I find that the 
calculated $\kappa$ of \gao is larger than that of \ino at all temperatures, which indicates that \gao can 
also exhibit peak $\kappa$ in excess of 1000 W/mK at low temperatures.  

I also calculated the thermal conductivity
of KTaO$_3$ ignoring the temperature-dependent softening of low-frequency phonons to investigate how the 
microscopic mechanisms for heat transport varies between binary and ternary oxides.  At high temperatures,
I find that the $\kappa$ of KTaO$_3$ and \gao are similar, reflecting the comparable magnitudes of their
acoustic phonon velocities.  However, the $\kappa$ of KTaO$_3$ increases less steeply at low temperatures 
compared to the binary oxides, and KTaO$_3$ has the lowest calculated $\kappa$ despite having acoustic phonon
velocities larger than that of $\beta$-In$_2$O$_3$.  I attribute this to the fact that the distribution of the
velocities of low-frequency acoustic phonons is more scattered in KTaO$_3$, and this results in enhanced 
momentum loss even during normal phonon-phonon scattering processes.  In the binary compounds, the acoustic
phonon branches are relatively less dispersive than in KTaO$_3$.  This results in a relatively narrower 
distribution of the acoustic phonon velocities at low frequencies in the binary oxides, which implies that
there is less momentum loss when the heat-carrying long-wavelength acoustic phonons scatter with each other. 
I think that the differences in acoustic phonon velocity distribution explains why the binary oxides have 
larger low-temperature $\kappa$ than the ternary oxides.

Finally, I also calculated the thermal diffusivity using the theoretically obtained thermal conductivity 
and heat capacity.  I find that all four compounds exhibit the expected $T^{-1}$ behavior at high temperatures.
I also used the calculated acoustic phonon velocities to computed the ratio $s$ of the average phonon scattering time
to Planckian time.  I find that $s$ obeys the recently found empirical lower bound of 1 even at 1000 K in 
all materials, which implies that the phonon-phonon collisions cannot happen faster than the Planckian time.

\section{acknowledgement}
I am grateful to Kamran Behnia for discussions and suggestions that motivated this work and for sharing 
experimental data.  The computational resources for this work was provided by by GENCI-CINES 
(grant 2019-A0070911099), the Swiss National Supercomputing Center (grant s820), and the European 
Research Council (grant ERC-319286 QMAC).

\end{document}